\documentclass{jltp}

\usepackage{multirow,amssymb,amsbsy,amsmath}
\usepackage{graphicx} 

\newcommand{\op}[1]{%
    \fontdimen12\textfont3=2pt\fontdimen12\scriptfont3=1.4pt%
    \!\null\mathop{\vphantom{#1}\smash{#1}}\limits_{\sim}\null\!}

\newcommand{\figref}[1]{Fig.~\protect\ref{#1}}
\newcommand{\fmref}[1]{(\protect\ref{#1})}

\newcommand {\mofe} {\{$\textrm{Mo}_{72}\textrm{Fe}_{30}$\}}
\newcommand{\pp}[2]{\frac{\partial \, {#1}}{\partial \, {#2}}}

\newcommand{\Bcrit}{B_{\text{c}}}

\newtheorem{k-rule}{k-rule}

\begin{document}

\title{Frustration effects in magnetic molecules}

\author{J\"urgen Schnack}
\address{Universit\"at Osnabr\"uck, Fachbereich Physik, D-49069 Osnabr\"uck, Germany}
\runninghead{J. Schnack}{Frustration effects in magnetic molecules}


\maketitle

\begin{abstract}
  Besides being a fascinating class of new materials, magnetic
  molecules provide the opportunity to study concepts of
  condensed matter physics in zero dimensions. This contribution
  will exemplify the impact of molecular magnetism on concepts
  of frustrated spin systems. We will discuss spin rings and the
  unexpected rules that govern their low-energy behavior.
  Rotational bands, which are experimentally observed in various
  molecular magnets, provide a useful, simplified framework for
  characterizing the energy spectrum, but there are also
  deviations thereof with far-reaching consequences. It will be
  shown that localized independent magnons on certain frustrated
  spin systems lead to giant magnetization jumps, a new
  macroscopic quantum effect. In addition a frustration-induced
  metamagnetic phase transitions will be discussed, which
  demonstrates that hysteresis can exist without anisotropy.
  Finally, it is demonstrated that frustrated magnetic molecules
  could give rise to an enhanced magnetocaloric effect.

PACS numbers: 75.50.Xx,75.10.Jm,75.40.Cx
\end{abstract}



\section{INTRODUCTION}
\label{sec-0}

Geometric frustration of interacting spin systems is the driving
force of a variety of fascinating phenomena in low-dimensional
magnetism.\cite{Gre:JMC01} In this context the term
\emph{frustration} describes a situation where in the ground
state of a classical spin system not all interactions can be
saturated simultaneously. A typical picture for such a situation
is a triangle of antiferromagnetically coupled spins, where
classically the spins are not in the typical up-down-up
configuration, but assume a ground state that is characterized
by a relative angle of $120^\circ$ between neighboring spins.
This special classical ground state characterizes several
frustrated spin systems, among them giant Keplerate
molecules,\cite{MLS:CPC01} the triangular lattice
antiferromagnet, and the kagome lattice antiferromagnet.

Throughout the article the spin systems are modeled by an
isotropic Heisenberg Hamiltonian augmented with a Zeeman term,
i.e.,
\begin{eqnarray}
\label{E-2-1}
\op{H}
&=&
-
\sum_{u, v}\;
J_{uv}\,
\op{\vec{s}}(u) \cdot \op{\vec{s}}(v)
+
g \mu_B B \op{S}_z
\ .
\end{eqnarray}
$\op{\vec{s}}(u)$ are the individual spin operators at sites
$u$, $\op{\vec{S}}$ is the total spin operator, and $\op{S}_z$
its $z$-component along the homogeneous magnetic field axis.
$J_{uv}$ are the matrix elements of the symmetric coupling
matrix. In the following we will consider only antiferromagnetic
couplings which are characterized by a negative value of
$J_{uv}$.

\section{GENERALIZED SIGN RULE FOR SPIN RINGS}
\label{sec-1}

Rigorous results on spin systems such as the sign rule of
Marshall and Peierls\cite{Mar:PRS55} and the famous theorems of
Lieb, Schultz, and Mattis \cite{LSM:AP61,LiM:JMP62} have
sharpened our understanding of magnetic phenomena. With the
advent of finite size antiferromagnetic spin rings the question
arose whether such general statements can also be made for odd
spin rings which are not decomposable into two sublattices, i.e.
not bipartite. A key quantity of interest is the shift quantum
number $k=0,\dots N-1$ associated with the cyclic shift symmetry
of the rings. The corresponding crystal momentum is then $2\pi
k/N$.  For rings with even $N$ (bipartite) one can explain the
shift quantum numbers for the relative ground states in
subspaces ${\mathcal H}(M)$ of total magnetic quantum number
$M$.\cite{Mar:PRS55,LSM:AP61,LiM:JMP62} In recent investigations
we could numerically verify, that even for frustrated rings with
odd $N$ astonishing regularities hold. Unifying the picture for
even and odd $N$, we find for the ground state without
exception:\cite{BHS:PRB03} \emph{The ground state
  belongs to the subspace ${\mathcal H}(S)$ with the smallest
  possible total spin quantum number $S$.  If $N\!\cdot\!s$ is
  integer, then the ground state is non-degenerate.  If
  $N\!\cdot\!s$ is half integer, then the ground state is
  fourfold degenerate.}

The sign rule of Marshall and Peierls can be generalized for all
subspaces ${\mathcal H}(M)$ with a given total magnetic quantum
number:\cite{BHS:PRB03} 
\begin{eqnarray}
\label{E-1-1}
\text{If}\;N\ne 3
&\text{ then }&\;
k
\equiv
\pm
a
\lceil
\frac{N}{2}
\rceil
\mod N
\ ,\quad
a=Ns-M
\ .
\end{eqnarray}
Moreover the degeneracy of the relative ground state is minimal.
Here $\lceil N/2\rceil$ denotes the smallest integer greater
than or equal to $N/2$. ``Minimal degeneracy" means that the
relative ground state in ${\mathcal H}(M)$ is twofold degenerate
if there are two different shift quantum numbers and
non-degenerate if $k=0$ mod $N$ or $k=N/2$ mod $N$, the latter
for even $N$.

The k-rule \fmref{E-1-1} is founded in a mathematically rigorous
way for $N$ even,\cite{Mar:PRS55,LSM:AP61,LiM:JMP62} $N=3$,
$a=0$, $a=1$, and in part for $a=2$.\cite{BHS:PRB03} For the
ground state with $N$ odd and $s=1/2$ the k-rule follows from the
Bethe ansatz.\cite{Kar94} An asymptotic proof can be formulated
for large enough $N$ for systems with an asymptotically finite
excitation gap (Haldane systems).\cite{BHS:PRB03} The k-rule
also holds for the exactly solvable $XY$-model with $s=1/2$.
For $N\!\cdot\!s$ being half integer field theory methods yield
that the ground state shift quantum number approaches $N/4$ for
large $N$.\cite{AGS:JPA89} Apart from these findings a rigorous
proof of the k-rule still remains a challenge.

\section{ROTATIONAL MODES AND MAGNETIZATION JUMPS}
\label{sec-2}

\begin{figure}[ht!]
\centering
\centerline{
\includegraphics[clip,width=90mm]{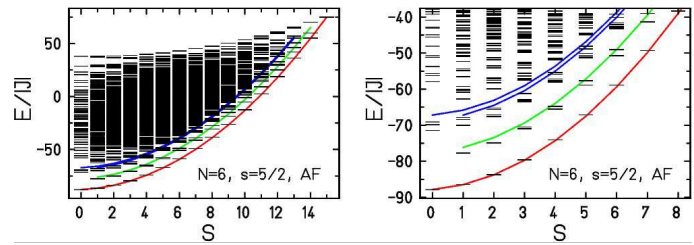}
}
\caption{Energy spectrum (l.h.s.) of a ring of six
  antiferromagnetically coupled spins $s=5/2$ and close-up view
  of the low-energy part (r.h.s.). The lowest rotational bands
  are highlighted by solid curves.}
\label{F-B}
\end{figure}
An antiferromagnet that can be decomposed into two sublattices
has as its lowest excitations the rotation of the N\'eel vector
as well as spin wave excitations.\cite{And:PR52} In finite size
systems these excitations are arranged in rotational (parabolic)
bands as shown in \figref{F-B} for a ring of six
antiferromagnetically coupled spins $s=5/2$. Such a behavior is
most pronounced for bipartite, i.e. unfrustrated
systems.\cite{ScL:PRB01,WGC:PRL03}

\begin{figure}[ht!]
\centering
\centerline{
\includegraphics[clip,width=90mm]{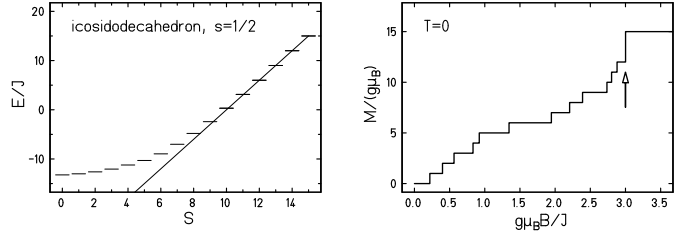}
}
\caption{L.h.s.: Minimal energies of the icosidodecahedron for
  $s=1/2$. The highest four levels fall on a straight line.
  R.h.s.: Resulting $(T=0)$-magnetization curve. The
  magnetization jump of $\Delta M=3$ is marked by an arrow.}
\label{F-C}
\end{figure}

Contrary to this behavior the minimal energies of certain
frustrated antiferromagnetic molecules of cuboctahedral and
icosidodecahedral structure depend linearly on total spin $S$
above a certain total spin.\cite{SSR:EPJB01} Such a dependence,
which is depicted on the l.h.s. of \figref{F-C}, results in an
unusually big jump to saturation as can be seen on the r.h.s. of
\figref{F-C}. Although first noticed for the Keplerate molecule
\mofe, such a behavior is quite common for a certain class of
frustrated spin systems such as the kagome or the pyrochlore
lattice.\cite{SHS:PRL02,RSH:JPCM03} The underlying reason is
that due to the special geometric frustration in such systems --
polygons are surrounded by triangles -- the relative ground
states in subspaces ${\mathcal H}(M)$ are for big enough $M$
given by product states of independent localized magnons.
Therefore, the energy scales linearly with the number of
independent magnons which in turn is linearly related to $M$ or
$S$.\cite{SHS:PRL02}

\section{HYSTERESIS WITHOUT ANISOTROPY}
\label{sec-3}

\begin{figure}[ht!]
\centering
\centerline{
\parbox[b]{60mm}{
\includegraphics[clip,width=49mm]{hysteresis-1.eps}
\\
\hspace*{6mm}\includegraphics[clip,width=45mm]{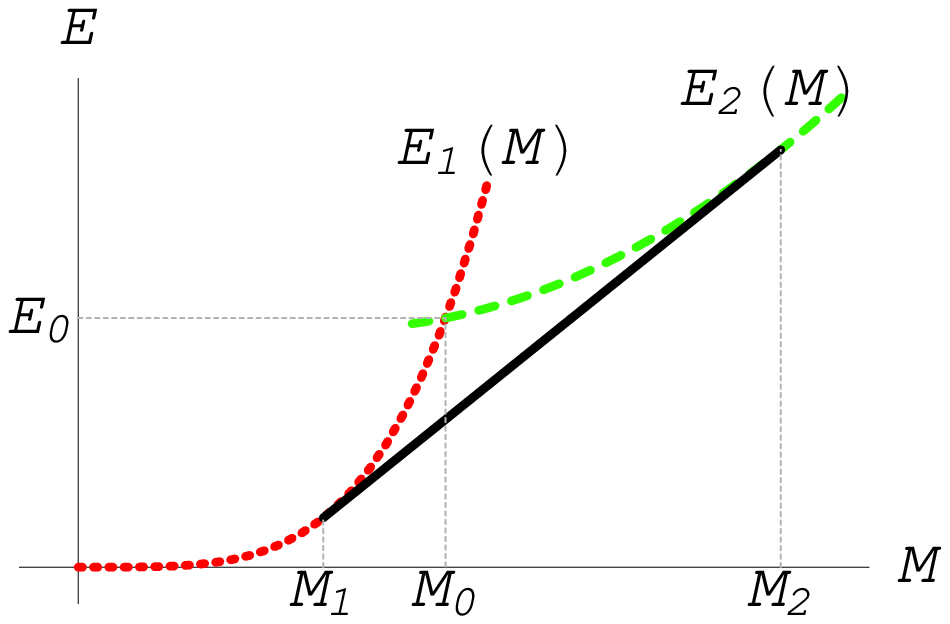}
}
\parbox[b]{60mm}{
\includegraphics[clip,width=55mm]{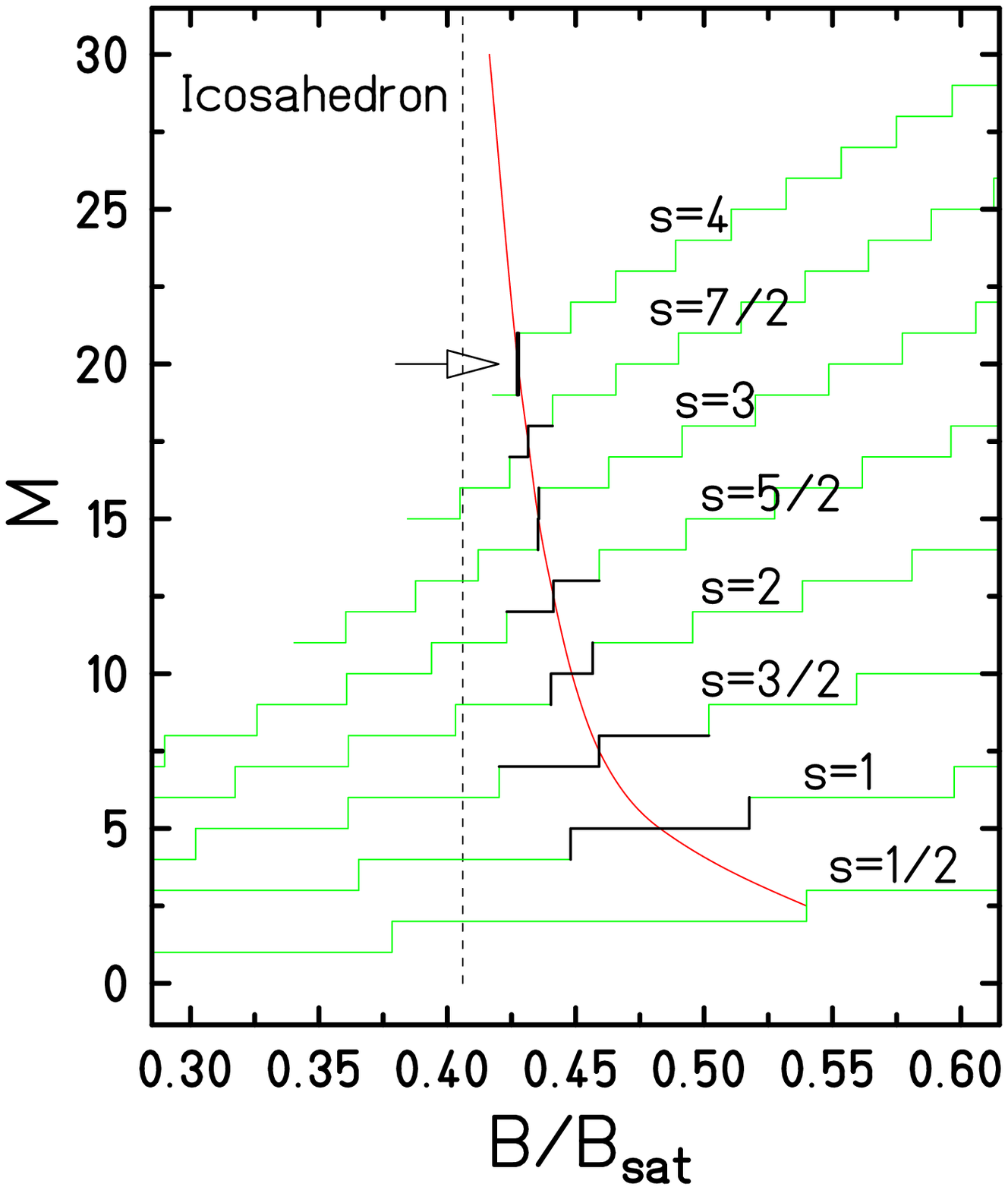}
}
}
\caption{L.h.s. (top): Hysteresis loop of the classical
  icosahedron. L.h.s. (bottom): Minimal energy curves of the
  classical icosahedron. R.h.s.: Magnetization curves of
  the quantum icosahedron for various $s$ at $(T=0)$.\cite{SSS:PRL05}}
\label{F-A}
\end{figure}

The observation of hysteresis effects in magnetic materials is
usually an outcome of their magnetic anisotropy.  In a recent
article we could report that magnetic hysteresis (\figref{F-A},
l.h.s., top) occurs in a spin system without any
anisotropy.\cite{SSS:PRL05} Specifically, we investigated an
icosahedron where classical spins mounted on the vertices are
coupled by antiferromagnetic isotropic nearest-neighbor
Heisenberg interaction giving rise to geometric frustration.  At
$T=0$ this system undergoes a first order metamagnetic phase
transition at a critical field $\Bcrit$ between two distinct
families of ground state configurations. The metastable phase of
the system is characterized by a temperature and field dependent
survival probability distribution.  Our exact classical
treatment shows that the abrupt transition at $T=0$ originates
in the intersection of two energy curves belonging to different
families of spin configurations that are ground states below and
above the critical field (\figref{F-A}, l.h.s., bottom). The
minimum of the two energy functions constitutes a non-convex
minimal energy function of the spin system and this gives rise
to a metamagnetic phase transition. We could also show that the
corresponding quantum spin system for sufficiently large spin
quantum number $s$ possesses a non-convex set of lowest energy
levels when plotted versus total spin.  \figref{F-A} (r.h.s.)
shows the $(T=0)$-magnetization curves for various $s$. The
magnetization plateaus of smallest width are highlighted on each
curve. At $s=4$ a magnetization jump of $\Delta M = 2$ occurs,
marked by the arrow. At $s=3$ a tiny plateau persists. The solid
curve shows that the field values that bisect the smallest
plateaus converge to the classical transition field (dashed
line).

\section{ENHANCES MAGNETOCALORIC EFFECT}
\label{sec-4}

The magnetocaloric effect, which consists in cooling or heating
of a magnetic system in a varying magnetic field, can assume
especially large values if the entropy $S(T,B)$ changes
drastically as a function of field according to
\begin{eqnarray}
\label{magmol-2-3}
\left(
\pp{T}{B}
\right)_S
&=&
-T
\frac{\left(\pp{S}{B}\right)_T}{C(T,B)}
\ .
\end{eqnarray}
This can for instance happen at phase transitions. In the
context of frustrated spin systems huge cooling rates
\fmref{magmol-2-3} can be achieved for certain topologies in the
vicinity of the saturation field due to the large (sometimes
even macroscopic) degeneracy of independent magnon
states.\cite{ZhH:JSM04,DeR:PRB04,SSR:05}

\begin{figure}[ht!]
\centering
\centerline{
\includegraphics[clip,width=45mm]{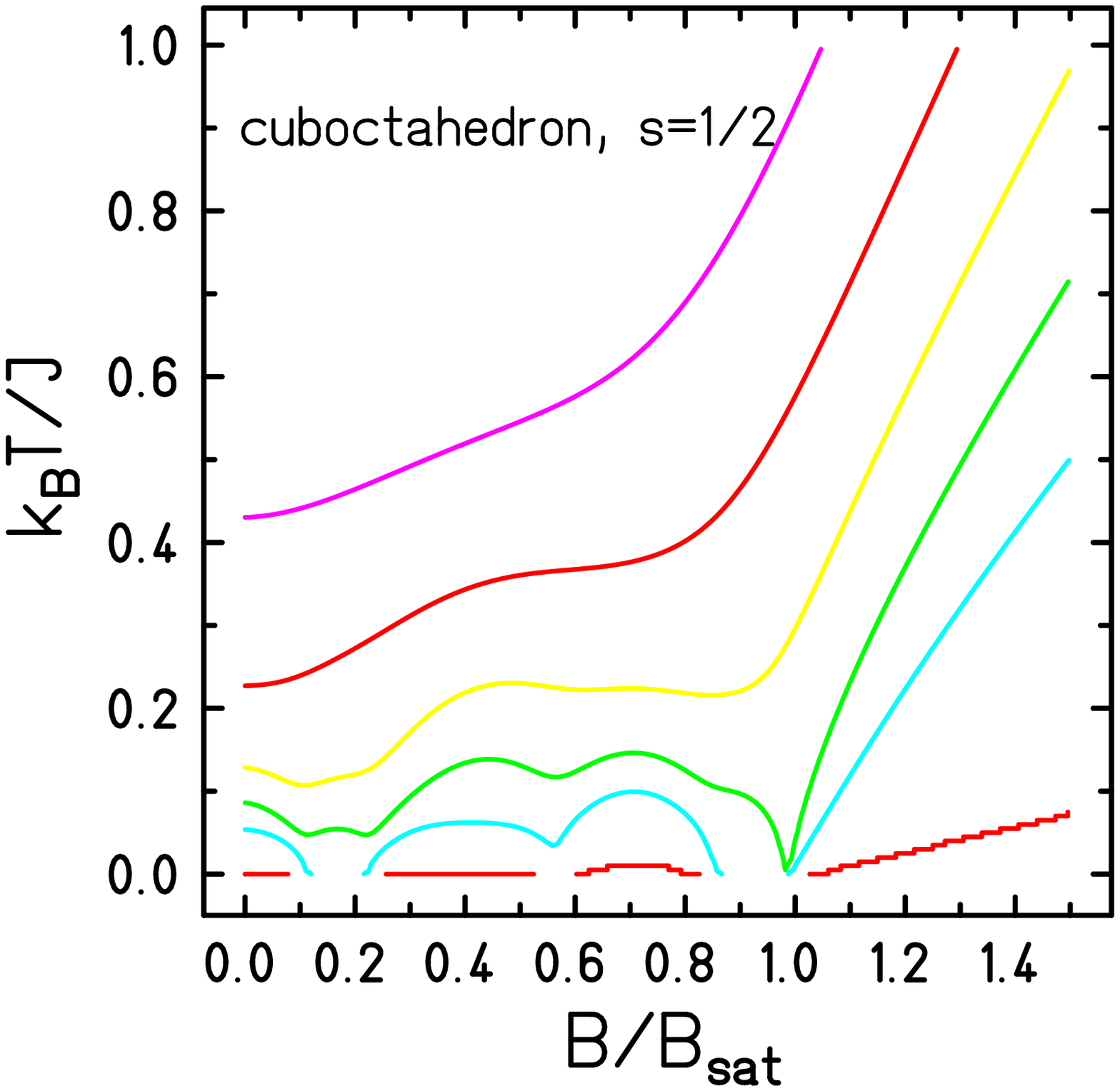}
$\quad$
\includegraphics[clip,width=45mm]{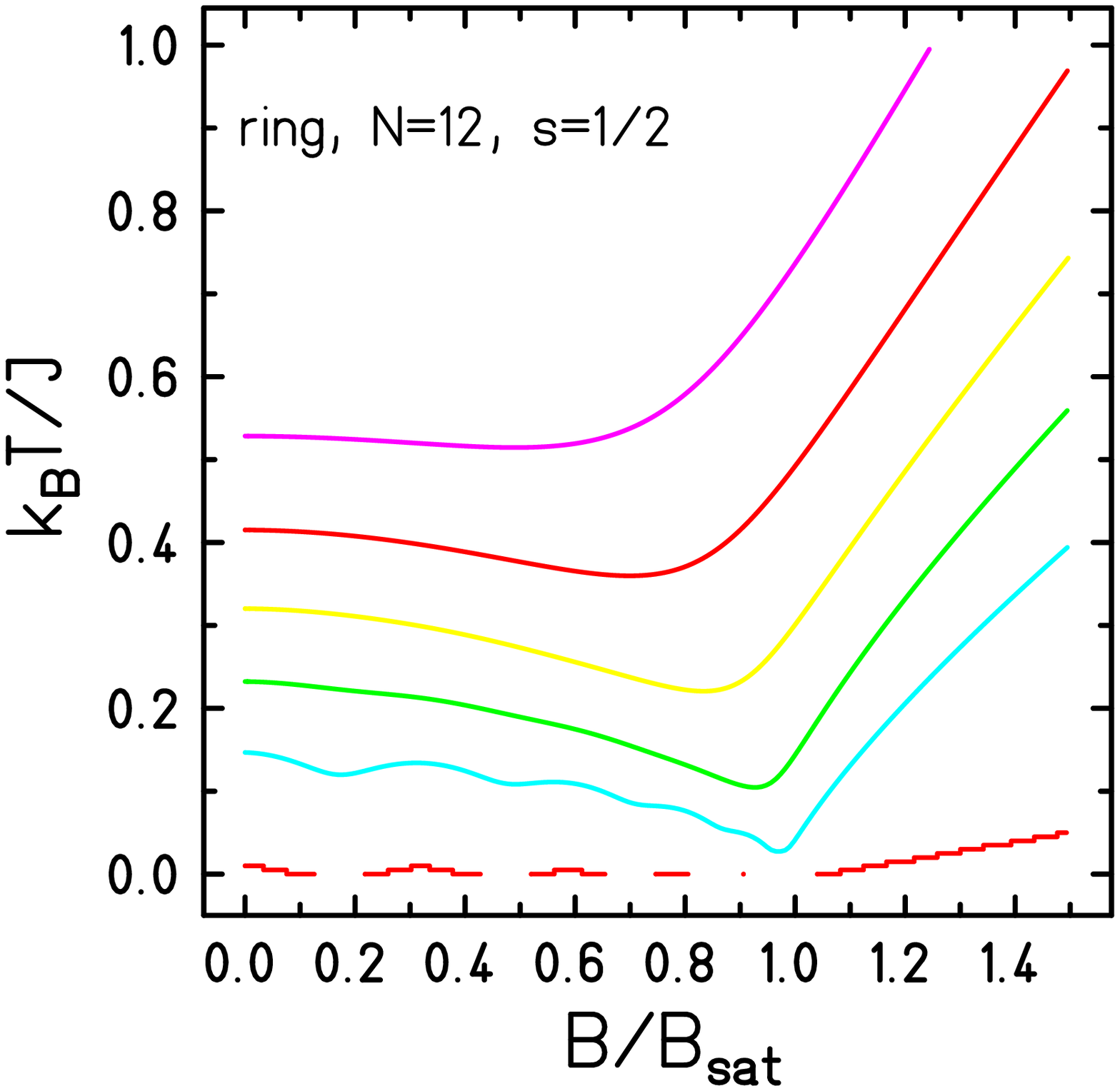}}
\caption{Isentropes of the cuboctahedron with $s=1/2$ (l.h.s.)
  as well as a spin ring with $N=12$ spins $s=1/2$ (r.h.s.).}
\label{F-D}
\end{figure}

\figref{F-D} shows the isentropes of a cuboctahedron with $s=1/2$ (l.h.s.)
as well as a spin ring with $N=12$ spins $s=1/2$ (r.h.s.). The
cuboctahedron hosts independent magnons and thus features an
unusually big magnetization jump to saturation whereas the spin
ring does not. Consequently the isentropes of the cuboctahedron
have a steeper slope above the saturation field, therefore the
cuboctahedron exhibits a larger cooling rate.

\section*{ACKNOWLEDGMENTS}

It is my pleasure to thank K.~B\"arwinkel, A.~Honecker,
P.~K\"ogerler, M.~Luban, J.~Richter, H.-J.~Schmidt, and
C.~Schr\"oder for the fruitful collaboration that produced so
many exciting results.


\begin{thebibliography}{10}

\bibitem{Gre:JMC01}
J. Greedan, J. Mater. Chem. {\bf 11},  37  (2001).

\bibitem{MLS:CPC01}
A. M\"uller {\it et~al.}, Chem. Phys. Chem. {\bf 2},  517  (2001).

\bibitem{Mar:PRS55}
W. Marshall, Proc. Royal. Soc. A (London) {\bf 232},  48  (1955).

\bibitem{LSM:AP61}
E.~H. Lieb, T. Schultz, and D.~C. Mattis, Ann. Phys. (N.Y.) {\bf 16},  407
  (1961).

\bibitem{LiM:JMP62}
E.~H. Lieb and D.~C. Mattis, J.~Math. Phys. {\bf 3},  749  (1962).

\bibitem{BHS:PRB03}
K. B\"arwinkel, P. Hage, H.-J. Schmidt, and J. Schnack, Phys. Rev. B {\bf 68},
  054422  (2003).

\bibitem{Kar94}
M. Karbach, Ph.D. thesis, Bergische Universit\"at - Gesamthochschule Wuppertal,
  1994.

\bibitem{AGS:JPA89}
I. Affleck, D. Gepner, H. Schulz, and T. Ziman, J. Phys. A {\bf 22},  511
  (1989).

\bibitem{And:PR52}
P.~W. Anderson, Phys. Rev. {\bf 86},  694  (1952).

\bibitem{ScL:PRB01}
J. Schnack and M. Luban, Phys. Rev. B {\bf 63},  014418  (2001).

\bibitem{WGC:PRL03}
O. Waldmann {\it et~al.}, Phys. Rev. Lett. {\bf 91},  237202  (2003).

\bibitem{SSR:EPJB01}
J. Schnack, H.-J. Schmidt, J. Richter, and J. Schulenburg, Eur. Phys. J. B {\bf
  24},  475  (2001).

\bibitem{SHS:PRL02}
J. Schulenburg {\it et~al.}, Phys. Rev. Lett. {\bf 88},  167207  (2002).

\bibitem{RSH:JPCM03}
J. Richter {\it et~al.}, J. Phys.: Condens. Matter {\bf 16},  S779  (2004).

\bibitem{SSS:PRL05}
C. Schr{\"o}der {\it et~al.}, Phys. Rev. Lett. {\bf 94},  207203  (2005).

\bibitem{ZhH:JSM04}
M.~E. Zhitomirsky and A. Honecker, J. Stat. Mech.: Theor. Exp. {\bf 2004},
  P07012  (2004).

\bibitem{DeR:PRB04}
O. Derzhko and J. Richter, Phys. Rev. B {\bf 70},  104415  (2004).

\bibitem{SSR:05}
J. Schnack, R. Schmidt, and J. Richter, in preparation.

\end{thebibliography}

\end{document}